\documentclass{iopconfser}

\usepackage{graphicx}

\usepackage{subfig}
\usepackage[english]{babel}
\usepackage{amsmath, amsthm, amssymb}
\usepackage{bbm}
\usepackage{url}
\usepackage{ulem}
\usepackage{tikz-cd}
\usepackage{braket}
\usepackage{path}
\usepackage{nicefrac}
\usepackage{amssymb}
 \usetikzlibrary{shapes,arrows}
\usepackage{amsmath}
\usepackage{amsthm}

\newtheorem{theorem}{Theorem}[section]
\newtheorem*{theorem*}{Theorem}

\theoremstyle{definition}
\newtheorem{definition}[theorem]{Definition}
\theoremstyle{corollary}

\theoremstyle{remark}

\theoremstyle{conclusion}

\begin{document}

\title{Subalgebra chains and nuclear physics: Commutant approach and construction of polynomial algebras }

\author{Rutwig Campoamor-Stursberg$^{1}$, Danilo Latini$^{2}$,  Ian Marquette$^{3}$,  Junze Zhang$^{4}$ and Yao-Zhong Zhang$^{4}$ }

\affil{$^1$Instituto de Matem\'{a}tica Interdisciplinar and Dpto. Geometr\'{i}a y Topolog\'{i}a, Universidad Complutense, UCM E-28040 Madrid, Spain }
\affil{$^2$ Dipartimento di Matematica “Federigo Enriques”, Università degli Studi di Milano, Via C. Saldini 50, 20133 Milano, Italy \& INFN Sezione di Milano, Via G. Celoria 16, 20133 Milano, Italy}
\affil{$^3$Department of Mathematical and Physical Sciences, La Trobe University, Bendigo VIC 3552, Australia  }
\affil{$^4$School of Mathematics and Physics, The University of Queensland, Brisbane QLD 4072, Australia  }

\email{rutwig@ucm.es, danilo.latini@unimi.it,  i.marquette@latrobe.edu.au,  junze.zhang@uq.net.au, yzz@maths.uq.edu.au }

\begin{abstract}
	In this paper, we review a new approach to study subalgebra chains $\mathfrak{g} \supset \mathfrak{g}'$ in the context of nuclear physics. This approach does not rely on explicit realizations as bosons or differential operators. We rely on the enveloping algebra, the notion of commutant $C_{U(\mathfrak{g})}(\mathfrak{g}^{\prime})$ and $\mathfrak{g}^{\prime}$-invariant polynomials. This approach builds on those $\mathfrak{g}^{\prime}$-invariant polynomials and finding the underlying finitely generated polynomial algebras. Those algebraic structures can then provide further information on sets of labeling operators. Another aspect of this method consists in exploiting the dual space and the symmetric algebra. Being independent of explicit realizations,  it endows the algebraic relations with a universal character. We review the chains associated with $\mathfrak{su}(3) \supset \mathfrak{so}(3)$, $\mathfrak{so}(5) \supset \mathfrak{su}(2) \times \mathfrak{u}(1)$,  $\mathfrak{su}(4) \supset \mathfrak{su}(2) \times \mathfrak{su}(2)$. Those chains are known as the Elliott, Seniority and Supermultiplet. We also provide new results and insights into the subalgebra chain $\mathfrak{so}(5) \supset \mathfrak{so}(3)$ of the Surfon model. For all chains, we present the related commutant, $\mathfrak{g}^{\prime}$-invariant polynomials and Poisson algebras.
\end{abstract}
\section{Introduction}
\label{int}

Over the years, different models in nuclear physics have attracted attention \cite{ia87,ia15}. They connect in different ways with Lie algebras, in particular in regard to dynamical symmetries, Casimir invariants, and subalgebra chains. Among the models that attracted attention are the Elliott \cite{Ell58a, Jud74}, Seniority \cite{Ahm70}, Supermultiplet \cite{Mos63b,Que76} , and Surfon models \cite{Mey85}. They are associated with the subalgebra chains $\mathfrak{su}(3) \supset \mathfrak{so}(3)$, $\mathfrak{so}(5) \supset \mathfrak{su}(2) \times \mathfrak{u}(1)$,  $\mathfrak{su}(4) \supset \mathfrak{su}(2) \times \mathfrak{su}(2)$ and $\mathfrak{so}(5) \supset \mathfrak{so}(3)$, respectively. The problem of missing labels also attracted attention. This occurs when the Casimir invariants associated with subalgebra chains do not provide enough labeling operators, and then other types of operators need to be constructed. Based on various methods for the search for Casimir invariants, methods for finding labeling operators were proposed in the context of their Poisson-Lie brackets and partial differential equations \cite{Cam07}. Later, approaches based on commutant (centralizer of subalgebra in the enveloping algebra of a Lie algebra) were introduced in the context of integrability and superintegrability \cite{Cam23a}. In this setting, polynomials that commute with the Cartan subalgebra were obtained for $A_n$. The corresponding polynomial algebras in the symmetric space were presented explicitly for the cases $n=2$ and $n=3$ \cite{Cam23a}.  Then, this approach was extended to the case of subalgebra chains related to nuclear physics and their missing labels. In this approach, $\mathfrak{g}^{\prime}$-invariant polynomials are constructed in the symmetric space using the related Poisson-Lie bracket \cite{Cam23c}. The homogeneous linearly independent and indecomposable polynomials are then classified, and a finite set characterizes all sets of labeling operators. Further insights are provided as this set contains more labeling operators than the minimal number. The operators cannot all Poisson commute with each other. They form a finitely generated polynomial algebra. This paper is devoted to reviewing the subalgebra chains and their related algebraic structures \cite{Cam23c,Cam25a,Cam25b} $\mathfrak{su}(3) \supset \mathfrak{so}(3)$, $\mathfrak{so}(5) \supset \mathfrak{su}(2) \times \mathfrak{u}(1)$,  $\mathfrak{su}(4) \supset \mathfrak{su}(2) \times \mathfrak{su}(2)$ and present new results concerning the $\mathfrak{so}(5) \supset \mathfrak{so}(3)$ chain. We believe that this summary of current research and new results on the Surfon chain will be of interest for researchers interested in algebraic structures and their applications in physics.

\section{Construction of commutant $C_{U(\mathfrak{g})}(\mathfrak{g}^{\prime})$ and $C_{S(\frak{g})}\left( \frak{g}^{\prime}\right)$ }
Let us recall some definitions and details on the construction of the commutant that will be exploited in later sections.
\begin{definition}
     Let $\mathfrak{g}:=\left\{X_1, \dots, X_n \, :\, [X_i,X_j]= \sum_{k=1}^nC_{ij}^k X_k\right\}$ be a $n$ dimensional semisimple or reductive Lie algebra over a field $\mathbb{F}$, and let us denote its universal enveloping algebra by $U(\mathfrak{g})$. Here $[\cdot,\cdot]: \mathfrak{g} \times \mathfrak{g} \rightarrow \mathfrak{g}$ is a commutator and $C_{ij}^k$  are the structure constants of $\mathfrak{g}$.  Let $\mathfrak{g}' \subset \mathfrak{g}$ be a $s$-dimensional subalgebra of $\mathfrak{g}$. The commutant $C_{U(\mathfrak{g})}(\mathfrak{g}^{\prime})$ of an $s$-dimensional subalgebra $\mathfrak{g}'$ is given by means of the condition  
\begin{equation}
	C_{U(\mathfrak{g})}(\mathfrak{g}^{\prime})=\left\{ P\in U(\mathfrak{g}): [X,P]=0,\quad \forall X\in\mathfrak{g}^{\prime}\right\} .\label{comm}
\end{equation} 
\end{definition}
\noindent The commutant is also denoted $C_{U(\mathfrak{g})}(\mathfrak{g}^{\prime})$ by $U(\mathfrak{g})^{\mathfrak{g}^{\prime}}.$ Let us provide more details on the construction of polynomial algebras in the dual space. Let $\mathfrak{g}^*$ be the dual space of $\mathfrak{g}$ with an ordered basis $\beta_{\mathfrak{g}^*} := \{e_1,\ldots,e_n\}$ such that $e_i(X_j) =\delta_{ij},$ where $\delta_{ij}$ is the Kronecker delta. We also assume that $\boldsymbol{x}=\{x_1, \ldots, x_n\}$ are the linear coordinates in $\mathfrak{g}^*$.  The algebra of smooth functions inherits the structure of a Poisson algebra that contains a subalgebra isomorphic to $\frak{g}$.  From this,  the linear Poisson bracket relations on $\mathfrak{g}^*$ take the form
\begin{equation}
	\{x_i,x_j\}= \sum_{k=1}^nC_{ij}^k x_k  \, ,
	\label{xcoord}
\end{equation} 
where $C^k_{ij}$ are the structure constants. Let $S(\mathfrak{g}) \cong \textsf{Pol}(\mathfrak{g}^*):=\mathbb{F}\left[ x_{1},\ldots ,x_{n}\right]$ be the symmetric algebra of $\mathfrak{g}$. Hence $S(\mathfrak{g}) $ can be identified as a finitely generated polynomial ring. By means of the Lie-Poisson bracket, in the coordinate form, we deduce that $\{\cdot,\cdot\} :S(\mathfrak{g}) \times S(\mathfrak{g}) \rightarrow S(\mathfrak{g})$ is given by
\begin{equation}
	\{p,q\}= \sum_{i,j,k =1}^n C_{ij}^k x_k \partial_{x_i} p\partial_{x_j} q  \, , \quad \text{ for any } p,q \in S(\mathfrak{g}).
	\label{LPB}
\end{equation}
We now construct the polynomial algebras from the subalgebras $\mathfrak{g}^\prime$ of $\mathfrak{g}$. 
\begin{definition}
\label{2.3} 
The kernel of the coadjoint action of $\mathfrak{g}^\prime$ on the  symmetric algebra $S(\mathfrak{g})$ is given by \begin{equation*}
	C_{S(\frak{g})}\left( \frak{g}^{\prime}\right) =\left\{ p\in S\left( 
	\frak{g}\right) :\left\{x ,p\right\} =0,\; x\in \frak{g}^{\prime *}\right\} .
\end{equation*} 
 \end{definition}
\noindent For convenience, we denote the Poisson centralizer $C_{S(\frak{g})}\left( \frak{g}^{\prime}\right)$ by $S(\mathfrak{g})^{\mathfrak{g}^\prime}$. Polynomials commuting with $\mathfrak{g}^{\prime}$ are determined as solutions of the equations in the following system of PDEs corresponding to the subalgebra generators $\left\{ x_u,p\right\}=0$:
\begin{equation}
	 \sum_{ l,k=1}^n C_{uk}^lx_l%
	\partial_{x_k} p=0,\; \text{ } 1\leq u\leq s=\dim \frak{g}^{\prime},
	\label{mlpa}
\end{equation}
where $(x_1, \dots, x_s)$ are coordinates in a dual basis of $\mathfrak{g}'^*$. We will then restrict ourselves to homogeneous polynomials of degree $k \geq 1$, as well as to their generic structure:
\begin{equation}
	p^{(k)}(\boldsymbol{x})=\sum_{a_1+\dots + a_n = k} \alpha_{a_1, \dots, a_n} \,x_1^{a_1} \dots x_n^{a_n}  \in S^k\left( \frak{g}\right) \,,
	\label{polynomials}
\end{equation} where $\alpha_{a_1,\ldots,a_n}$  are constants.  For embedding chains, elements in the Poisson centralizer $S(\mathfrak{g})^{\mathfrak{g}^\prime}$ correspond to polynomials of the form $p^{(k)}(\boldsymbol{x})$ that Poisson commute with the corresponding linear coordinates associated to the generators of the subalgebra $\mathfrak{g}'$ w.r.t. the Lie-Poisson bracket \eqref{LPB} i.e. $\left\{x_u, p^{(k)}(\boldsymbol{x}) \right\}=0$. All degree $k$ $\mathfrak{g}'$-invariant polynomials are denoted $\boldsymbol{p}^{(k)}:=\left\{p_1^{(1)}(\boldsymbol{x}), \ldots, p_{m_k}^{(1)}(\boldsymbol{x})\right\}$. We restrict the set to all linearly independent and indecomposable $\mathfrak{g}'$-invariant polynomials by iterating from lower order to maximal order $\zeta$ and filtering by removing decomposable polynomials from the knowledge of the lower-degree ones. The finitely generated polynomial (Poisson) algebras are obtained from the set $\boldsymbol{Q}^{[\zeta]}$. We aim to close the Poisson-Lie bracket in $\textbf{Alg} \left\langle \boldsymbol{Q}^{[\zeta]} \right\rangle $ by 

\begin{equation}
	\left\{p_{i_1}^{(h_{i_1})} , p_{i_2}^{(h_{i_2})}\right\}  =  \sum_{k_1+\ldots + k_r =h_{i_1} + h_{i_2}-1} \beta^{s_1,\ldots,s_r}_{i_1,i_2} p^{(k_1)}_{s_1} \cdots p^{(k_r)}_{s_r}  . \, 
	\label{polrel}
\end{equation} Here $\beta^{s_1,\ldots,s_r}_{i_1,i_2}$ are the coefficients, $h_{i_1},h_{i_2},k_1,\ldots,k_r \leq \zeta$ and $ 1  \leq s_1,\ldots,s_r \leq m_{[\zeta]}$. Since the generators in $\boldsymbol{Q}^{[\zeta]}$ are not functionally independent, to ensure $\eqref{polrel}$ is indeed a Poisson bracket, further polynomial relations need to be quotiented out. 

It is known that for a semisimple Lie algebra $\mathfrak{g}$ of rank $\ell$, the associated root system implies that states the well-known $i_0=\frac{1}{2}(\dim \frak{g}+\ell)$ labels as the size of a maximal commutative subalgebra of $U(\mathfrak{g})$. Alternatively, in physical applications, often some nonabelian subalgebra $\frak{g}^{\prime}$ (of rank $\ell^{\prime}$) is used to label the basis states and provides $\rho_0=\frac{1}{2}(\dim \frak{g}^{\prime}+\ell^{\prime}))-\ell_{0}$ labels, with $\ell_{0}$ being the number of algebraically independent common Casimirs, i.e., the rank of the algebra $\mathcal{Z}(U(\mathfrak{g})) \cap \mathcal{Z}(U(\mathfrak{g}'))$ and $\frak{g}^{\prime}$ and thus $n_0=\frac{1}{2}\left( \dim\frak{g}-\ell -\dim\frak{g}^{\prime}-\ell^{\prime})\right)+\ell_{0}$ additional commuting operators are needed. These operators must commute with the subalgebra $\frak{g}^{\prime}$ and hence belong to elements in the centralizer $C_{U(\frak{g})}\left(\frak{g}^{\prime}\right)$. Taking into account that the $i_0$ labeling operators must also commute with each other and span the Abelian subalgebra of $C_{U(\frak{g})}\left(\frak{g}^{\prime}\right)$. 
\section{The chain $\mathfrak{su}(3) \supset \mathfrak{so}(3)$: Elliott model}  
\label{sec3.1}

The first example we discuss concerns the chain associated with one of the most studied missing label problems,  the so-called Elliott chain $\mathfrak{su}(3) \supset \mathfrak{so}(3)$,  relevant to the study of the Elliott model in Nuclear Physics. In this chain, the $\mathfrak{so}(3)$ subalgebra is spanned by the three orbital angular momentum operators $\boldsymbol{L}=(L_1, L_2, L_3)$, while the remaining five generators are the components of a rank-two symmetric traceless tensor. We define the Lie-Poisson bracket \eqref{LPB} in terms of the linear coordinates: $\boldsymbol{x}:=\{x_1,x_2,x_3,x_4,x_5,x_6,x_7,x_8\}$ $\equiv$ $\{\ell_1, \ell_2, \ell_3, t_{11},t_{12},t_{13}, t_{22}, t_{23}\}$. In the Poisson setting, the commutators translate to the following ones:
\begin{align}
\{\ell_{i},\ell_{j}\}&= {\rm i} \epsilon_{ijk}\ell_k \, , \hskip 1cm
\{\ell_{i},t_{jk}\}= {\rm i} \epsilon_{ijl} t_{kl} + {\rm i} \epsilon_{ikl} t_{jl} \nonumber \\
\{t_{ij}, t_{kl}\}&= {\rm i} (\delta_{ki} \epsilon_{jlm} + \delta_{li} \epsilon_{jkm}+ \delta_{kj} \epsilon_{ilm} + \delta_{lj} \epsilon_{ikm} ) \ell_m \label{ee6} \, .
\end{align}
\noindent We now proceed with our systematic procedure to find the classical (unsymmetrized) elements defining the commutant of the rotation subalgebra corresponding to the coordinates $\boldsymbol{x}':=\{x_1, x_2, x_3\}\equiv\{\ell_1, \ell_2, \ell_3\}$.
Given the set of elements $\boldsymbol{x}=\{\ell_{1}, \ell_{2},\ell_{3},t_{11}, t_{12}, t_{13}, t_{22}, t_{23}\}$, the expansion (\ref{polynomials}) at order $n$ is given by 
\begin{equation}
p^{(n)}(\boldsymbol{x})=\sum_{a_1+\dots + a_8= n} \alpha_{a_1, \dots, a_8} \,x_1^{a_1} \dots x_8^{a_8} 
\label{eq:exp1}
\end{equation}
with the commutant constraint $\{\ell_k, p^{(n)}(\boldsymbol{x})\}=0,\quad k=1,2,3$. The construction terminates at degree $N=6$. The polynomial algebra closes in terms of the following seven elements $\{b_1, b_2, C_1, C_2, D_1, F_1\}$ with the following grading $\mathcal{G}\left(b_1\right)= (2,0)$, $\mathcal{G}\left( b_2 \right)= (0,2)$, $\mathcal{G}\left( C_1 \right)= (2,1)$,  $\mathcal{G}\left( C_2 \right)= (0,3)$,
  $\mathcal{G}\left( D_1 \right)= (2,2)$,  $\mathcal{G}\left( F_1 \right)= (3,3)$. 
The basis formed by the following six elements (the seventh having been discarded due $c_1:=b_1$, $c_2:=b_2$, $c_3:= \frac{1}{2} C_1+ C_2$, $\mathcal{A}:=\frac{1}{2} C_1-C_2$, $\mathcal{B}:=D_1$,  $\mathcal{C}:=-4 {\rm i} F_1$. Different notations have been used for the set of indecomposable \cite{Cam23c} $\{c_1, c_2, c_3, \mathcal{A},\mathcal{B}, \mathcal{C}\}:=\{p_1^{(2)}, p_2^{(2)}, p_1^{(3)}, p_2^{(3)}, p_1^{(4)}, p_1^{(6)}\}$. We then aim to get explicitly the structure constants as follows
\begin{equation}
\left\{p_{m_1}^{(i)},p_{m_2}^{(j)}\right\}= \sum_{2\alpha_1 + \ldots + 6\alpha_{6}=  i +j -1 }  \gamma^{\alpha_1,\ldots,\alpha_6}_{m_1,m_2}(p_1^{(2)})^{\alpha_1} (p_2^{(2)})^{\alpha_2} (p_1^{(3)})^{\alpha_3} ( p_2^{(3)})^{\alpha_4} (p_1^{(4)})^{\alpha_5} (p_1^{(6)})^{\alpha_6}.
\end{equation} Let us point out, the notation of this previous equation is different from (\ref{polrel}). The three-generator polynomial cubic Poisson algebra is obtained below:
\begin{equation}\label{eqc1}
\begin{split}
\{\mathcal{A},\mathcal{B}\}&=\mathcal{C} ,\quad
\{\mathcal{A},\mathcal{C}\}=\alpha \mathcal{A}^2+\beta \mathcal{B}^2 +\delta \mathcal{A}+\epsilon \mathcal{B} +\zeta  \\
\{\mathcal{B},\mathcal{C}\}&=\lambda \mathcal{A}^3+\mu \mathcal{A}^2+2\xi \mathcal{A} \mathcal{B}+\rho \mathcal{A}+\sigma \mathcal{B} +\chi  
\end{split}
\end{equation}
\begin{equation}\label{clst1}
\begin{split}
\alpha & =-\xi=-8(3 c_1+c_2) \, , \quad \beta=24 \, , \quad \delta=-\sigma=-16 c_2 c_3 \, , \quad \epsilon=-16 c_1 c_2 \, , \quad \zeta=8 c_3^2(3 c_1-c_2),\\
 \lambda & =-32,\quad \mu=-48 c_3,\quad \rho=-16 c_1(c_1+c_2)^2,\quad \chi= 16 c_3 (c_1(c_1^2 -  c_2^2) + c_3^2) .
\end{split}
\end{equation}

\section{The chain $\mathfrak{so}(5) \supset \mathfrak{su}(2) \times \mathfrak{u}(1)$}
\label{sec3.2}

\noindent The reduction chain $\mathfrak{so}(5) \supset \mathfrak{su}(2)$ also constitutes a chain that has found various applications in nuclear physics, such as, e.g., the surface quadrupole vibrations or the nuclear seniority model. For this reduction chain, the system (\ref{mlpa}) has six (functionally) independent solutions. Hence, a polynomial algebra associated with this reduction chain must possess at least six generators. Four of these solutions correspond to the Casimir operators of $\mathfrak{so}(5)$ and $\mathfrak{su}(2) \times \mathfrak{u}(1)$, respectively, so two additional functions are necessary.  The ten-dimensional Lie algebra $\mathfrak{so}(5)$ is given in terms of the basis $\{S_-, T_-, U_-, V_-, S_+, T_+, U_+,V_+,U_3, V_3\}$ and the $\mathfrak{su}(2) \times \mathfrak{u} (1)$ subalgebra is spanned by the basis $\{U_-, U_+, U_3, V_3\}$. With respect to the commutative coordinates: $\boldsymbol{x}:=\{x_1,x_2,x_3,x_4,x_5,x_6,x_7,x_8, x_9,x_{10}\} \equiv\{s_-,t_-,u_-,v_-,s_+,t_+,u_+,v_+,u_3, v_3\},$ the subalgebra is spanned by: $\boldsymbol{x}':=\{x_3, x_7, x_9,x_{10}\}\equiv\{u_-,u_+, u_3, v_3\}, $
and the commutant is obtained from the system:
\begin{equation}
\{u_\pm, p^{(n)}(\boldsymbol{x})\} =0, \quad \{u_3,\, p^{(n)}(\boldsymbol{x})\},\quad 
\{v_3, \hskip 0.1cm p^{(n)}(\boldsymbol{x})\}=0. 
\label{commso5}
\end{equation}
 Following the implementation of the procedure gives rise to the following nine polynomials up to degree six. It can be clearly shown that for all monomials of the polynomial, the sum of the eigenvalues relative to $u_3$ and $v_3$ is zero. Additionally,
\[     \mathcal{G}\left(a_1\right)= (1,0), \quad         \mathcal{G}\left(b_1\right)= (2,0), \quad    \mathcal{G}\left(b_2\right)= (0,2), \quad         \mathcal{G}\left(C_1\right)= (1,2) \]
\[        \mathcal{G}\left(D_1\right)= (0,4), \quad  \mathcal{G}\left(D_2\right)= (0,4),  \quad \mathcal{G}\left(D_4\right)= (2,2),  \quad \mathcal{G}\left(F_1\right)= (2,4) ,\quad 
\mathcal{G}\left(F_2\right)= (2,4). \] Here $\mathcal{G}$ is the grading for a polynomial generator \cite{Cam25a}. The grading provides important constraints on the admissible monomials in the Poisson bracket relations. Higher-order polynomials do not provide additional independent elements. The polynomials $a_1$ and $b_1$ correspond to the Casimir operators of the subalgebra, that $a_1^2+b_1+b_2$ corresponds to the quadratic Casimir operator of $\mathfrak{so}(5)$, and the fourth-order invariant is given by  $D_1+D_2+ 4a_1^2(b_1+b_2)+4b_1b_2+ 2b_1^2+ 2a_1^4$. There are functional relations and we can restrict \cite{Cam23c} to indecomposable polynomials $c_1:=a_1$, $c_2:=b_1$, $c_3:= b_2$, $c_4=D_1+D_3+4 a_1 C_1$,  $\mathcal{A}:=C_1$, $\mathcal{B}:=D_1$ and $\mathcal{C}:=2(F_1 -F_2)$. The set can also be denoted as
$\{c_1,c_2,c_3,c_4,\mathcal{A}, \mathcal{B},\mathcal{C}\} =\{p_1^{(1)},p_1^{(2)} , p_2^{(2)} , p_1^{(4)}, p_1^{(3)},p_1^{(4)}, p_1^{(6)}\}$. The Poisson algebra then takes the form as follows:
\begin{equation}
\{p_{m_1}^{(i)},p_{m_2}^{(j)}\}= \sum_{\alpha_1 + \ldots + 6 \alpha_{7}= i +j -1} \gamma^{\alpha_1,\ldots, \alpha_7}_{m_1,m_2}(p_1^{(1)})^{\alpha_1} (p_1^{(2)})^{\alpha_2} (p_2^{(2)})^{\alpha_3}  (p_1^{(4)})^{\alpha_4} ( p_1^{(3)})^{\alpha_5} (p_2^{(4)})^{\alpha_6} (p_1^{(6)})^{\alpha_7},
\end{equation}  and explicitly close as a cubic algebra similar to the Elliott chain $\mathfrak{su}(3) \supset \mathfrak{so}(3)$ but with structure constants given by
\begin{equation}
\begin{split}
\alpha&=-\xi= 8 (4 c_1^2 - 2 c_2 - c_3), \quad \beta=6 \, , \quad \gamma=-\nu=16 c_1 \, , \quad \delta=-\sigma=- 
16  c_1 (c_3^2 + c_4),\\
 \epsilon&=4 (4 c_2^2 - c_3^2 - 2 c_4), \quad \zeta=4 c_3^2( c_4-6 c_2^2) + 2c_4^2, \quad \lambda=-32, \quad \mu=96c_1 c_3,\\
 \rho&=16 c_3 (4 ( c_1^2 - c_2) c_3 - c_4), \quad \chi=  -16 c_1 c_3^2 c_4,
\end{split}
\end{equation}
where $c_i$ ($i=1,2,3,4$) are central elements. See \cite{Cam23c} for details.
\section{The reduction chain $\mathfrak{su}(4) \supset \mathfrak{su}(2) \times \mathfrak{su}(2)$}
\label{subsec3.1}

The state labeling problem for the reduction chain
$\mathfrak{su}(4) \supset \mathfrak{su}(2) \times \mathfrak{su}(2)$, has been analyzed by various authors using different techniques. As this problem is computationally difficult, the polynomial and algebraic structure are quite involved. In this section, we will summarize results. The fifteen-dimensional Lie algebra $\mathfrak{su}(4)$ is given in terms of the basis: $\{S_1, S_2, S_3, T_1, T_2, T_3, Q_{11}, Q_{12}, Q_{13}, Q_{21}, Q_{22}, Q_{23}, Q_{31}, Q_{32}, Q_{33}\} $ and the $\mathfrak{su}(2) \times \mathfrak{su}(2)$ subalgebra is spanned by the six generators $\{S_1, S_2, S_3, T_1, T_2, T_3\}$. Under this basis, the Lie algebra $\mathfrak{su}(4)$ admits the (vector space) decomposition $ \mathfrak{su}(4) = \mathfrak{g}_1 \oplus \mathfrak{g}_2 \oplus \mathfrak{g}_3$ with $\mathfrak{g}_1 = \mathrm{span} \{S_1,S_2,S_3\}$, $\mathfrak{g}_2 = \mathrm{span} \{T_1,T_2,T_3\}$  and  $\mathfrak{g}_3 = \mathrm{span} \{Q_{11},\ldots,Q_{33}\}$. In this case, we consider the following linear coordinates:  $  \boldsymbol{x}:=  \{x_1, x_2, x_3, x_4, x_5, x_6,x_7,x_8,x_9,x_{10},x_{11},x_{12}, x_{13}, x_{14}, x_{15}\}$ $  =  \{s_1,s_2,s_3,t_1,t_2,$ $ t_3,q_{11},q_{12},q_{13}, q_{21},q_{22},q_{23}, q_{31}, q_{32},q_{33}\}. $ In the commutative setting, the brackets of coordinates \eqref{xcoord} explicitly read:
\begin{equation}
\begin{split}
\{s_i, s_j\}&={\rm i} \epsilon_{ijk}s_k \qquad \{t_\alpha, t_\beta\}={\rm i} \epsilon_{\alpha \beta \gamma}t_\gamma \qquad \{s_i, t_\alpha\}=0 \\
\{s_i,q_{j\alpha}\}&={\rm i} \epsilon_{ijk} q_{k \alpha} \quad \,\, \{t_\alpha, q_{i\beta}\}={\rm i} \epsilon_{\alpha \beta \gamma} q_{i \gamma} \quad \{q_{i \alpha}, q_{j \beta}\}=\frac{{\rm i}}{4}(\delta_{\alpha \beta} \epsilon_{ijk}s_k+\delta_{ij}\epsilon_{\alpha \beta \gamma}t_\gamma) \, .
\label{classicalrels}
\end{split}
\end{equation}
Here $\delta_{ij}$ is the Kronecker delta and $\epsilon_{ijk}$ is the Levi-Civita symbol. These relations will serve as the building blocks for deriving the polynomial algebra relations. We will present the classical elements defining $\mathcal{Q}_{\mathfrak{su}(4) }(d) := \left(  \textbf{Alg} \left\langle \boldsymbol{Q}^{[9]}\right\rangle,\{\cdot,\cdot\}\right)$. Here $d$ is the degree of the polynomial algebra. The general homogeneous polynomial at degree $k$ is given by 
\begin{equation}
p^{(k)}(\boldsymbol{x})=\sum_{a_1+\ldots + a_{15}= {k}} \Gamma_{a_1, \dots, a_{15}} \,x_1^{a_1} \cdots x_{15}^{a_{15}}  
\label{eq:exp1}
\end{equation}
with the commutant constraint: $\{s_i, p^{(k)}(\boldsymbol{x})\}=0$, $i=1,2,3$, $\{t_\alpha, p^{(k)}(\boldsymbol{x})\}=0$, $\alpha=1,2,3.$ The list of polynomials is $\boldsymbol{Q}^{[9]} =  \{b_1, b_2, b_3, c_1, C_2, d_1, D_2, D_3, D_4, E_1, F_1, F_2, F_3, F_4, G_1, G_2, H_1, H_2, I_1, I_2\}$. At this point, a crucial observation that ensures the correctness of our results is that the twenty polynomials we have obtained are directly connected to the ones reported in \cite{Que76}. Let us give the grading of the polynomials of $\boldsymbol{Q}^{[9]}$: 
\begin{equation}
\begin{split}
b_1&=C^{(200)}  \, , \quad b_2=C^{(020)} \, , \quad b_3=C^{(002)} \, , \quad c_1=C^{(111)}-\frac{2}{3}C^{(003)} \, , \quad C_2=C^{(111)}\\
d_1&=C^{(202)}+C^{(022)}-C^{(112)}-2 C^{(004)} \, ,\quad D_2=\frac{1}{2}C^{(112)} \, , \quad D_3=C^{(022)} \, ,\quad D_4=C^{(004)} \, , \\
 E_1&=C^{(113)} \, ,\quad  F_1=C^{(213)} \, , \quad F_2=2 C^{(204)}-C^{(200)}C^{(004)} \, , \quad F_3=C^{(123)} \, , \\
 F_4&=2 C^{(024)}-C^{(020)}C^{(004)} \, , \quad G_1=C^{(214)} \, , \quad  G_2=C^{(124)} \, , \quad H_1= C^{(215)} \, , \quad H_2=C^{(125)} \, , \\
 I_1&=-8C^{(036)} \, , \quad I_2=-4\bigl(C^{(306)}+C^{(002)}C^{(214)}+\frac{1}{6}C^{(003)} C^{(213)}\bigl) \, .
\end{split} 
\label{poldanilo}
\end{equation}
Using simplification $\bar{F}_2:=C^{(204)}$, $\bar{F}_4:= C^{(024)}$ and $\bar{I}_2:=C^{(306)}$. All others remain the same but now denoted uniformly. The next step consists in constructing the polynomial algebra generated by the following set composed of twenty polynomials up to degree nine (five central elements and fifteen generators). Hence $\textbf{Alg} \left\langle \mathcal{P} \right\rangle $ generates a polynomial algebra with $\dim_{FL} \textbf{Alg} \left\langle \mathcal{P} \right\rangle = 20$. In the next section, we will show that $\textbf{Alg} \left\langle \mathcal{P} \right\rangle$ is closed under the Poisson bracket $\{\cdot,\cdot\}$. Additionally, we should highlight that the two missing label operators, which Moshinsky and Nagel discussed in \cite{Mos63b} and represented by two commuting $\mathfrak{su}(2) \times \mathfrak{su}(2)$ scalars within the enveloping algebra of $\mathfrak{su}(4)$, can be expressed as follows:
\begin{equation}
\Omega=C^{(111)} =\bar{C}_2\, , \qquad \Phi=C^{(202)}+C^{(022)}-C^{(112)}=\bar{d}_1+2\bar{D}_4 \, .
\end{equation}
The structure of the polynomial algebra is generated by the set $\mathcal{P}$. In detail, with the five elements $\{\bar{b}_1, \bar{b}_2, \bar{b}_3, \bar{c}_1,$ $ \bar{d}_1\}$, the generating set is given by: $$\boldsymbol{Q}^{[9]} =  \left\{ p_1^{(2)}, p_2^{(2)}, p_3^{(2)}, p_1^{(3)}, p_2^{(3)}, p_1^{(4)}, p_2^{(4)}, p_3^{(4)}, p_4^{(4)}, p_1^{(5)}, p_1^{(6)}, p_2^{(6)}, p_3^{(6)}, p_4^{(6)},  p_1^{(7)} ,  p_2^{(7)}, p_1^{(8)}, p_2^{(8)}, p_1^{(9)}, p_2^{(9)}  \right\}$$ with the following closed relations: \begin{equation}
\left\{p_{m_1}^{(i)},p_{m_2}^{(j)}\right\}= \sum_{2\alpha_1 + \ldots + 9 \alpha_{20}= i +j -1}  \gamma^{\alpha_1,\ldots,\alpha_{20}}_{m_1,m_2} ( p_1^{(2)}  )^{\alpha_1} \quad \times \ldots \times \quad  (   p_2^{(9)}  )^{\alpha_{20}}
\end{equation} 
 When considering expansions, we formally denote the sets with elements of degree two, three, four, and six as $\mathbf{B}$, $\mathbf{C}$, $\mathbf{D}$, and $\mathbf{F}$ respectively. We get the following for the lowest degree:
\begin{equation}
\{C_2, D_2\}=-{\rm i} (F_1+F_3) \, , \qquad \{C_2, D_3\}= -2 {\rm i} F_3 \, , \qquad \{C_2, D_4\}=0 \, .
\end{equation}
and for the highest degree Poisson relations for which we provide the highest-order terms \begin{align}
\{\bar{I}_1, \bar{I}_2\}&= -\frac{7}{8} i b_2 b_3 D_3 I_2-\frac{7}{16} i b_2 b_3 D_2 I_2+\frac{3}{4} i D_3^2 I_2 + l.o.t
\end{align}
Here higher and lower-order terms refer to lexicographic ordering (they all have the same overall degree).

\section{Surfon model from Lie algebra chain $\mathfrak{so}(5) \supset \mathfrak{so}(3)$}
We now consider the chain of $\mathfrak{so}(5) \supset \mathfrak{so}(3)$ corresponding to the Surfon model. We will present results on the classification of polynomials of the commutant and the polynomial algebra: \begin{align}
 [L_0, L_{-1}]&= -L_{-1},\quad [L_0, L_1]= L_1,\quad [L_1, L_{-1}]= 2 \ell_0, \quad [L_0, Q_3]= 3 Q_3,\quad [L_0, Q_2]= 2 Q_2,  \\
  [L_0, Q_1]&= Q_1, [L_0, Q_0]=0,\quad [L_0, Q_{-1}]= -Q_{-1},\quad [L_0, Q_{-2}]= -2 Q_{-2},  [L_0, Q_{-3}]= -3 Q_{-3} \nonumber \\
 [L_1, Q_3]&=0,\quad [L_1, Q_2]= 6 Q_3,\quad [L_1, Q_1]= Q_2,\quad [L_1, Q_0]= 2 Q_1,\quad [L_1, Q_{-1}]= 6 Q_0, \nonumber \\
 [L_1, Q_{-2}]&= 10 Q_{-1},\quad [\ell_1, Q_{-3}]= Q_{-2},\quad [L_{-1}, Q_3]= Q_2, ,\quad [ L_{-1}, Q_2]= 10 Q_1,\quad [L_{-1}, Q_1]= 6 Q_0, \nonumber \\
 [L_{-1}, Q_0]&= 2 Q_{-1}, \quad [ L_{-1}, Q_{-1}]= Q_{-2},\quad [ L_{-1}, Q_{-2}]= 6 Q_{-3},\quad [ L_{-1}, Q_{-3}]=0,\quad [ Q_3, Q_2]=0  \nonumber \\
 [ Q_3, Q_2]&=0,\quad [Q_3, Q_1]=0,\quad [Q_3, Q_0]= Q_3, \quad [Q_3, Q_{-1}]= Q_2,\quad [ Q_3, Q_{-2}]= 10 Q_1 + 15 L_1, \nonumber \\
 [ Q_3, Q_{-3}]&= 5 Q_0 - 15 L_0, \quad [ Q_2, Q_1]= -6 Q_3,\quad [Q_2, Q_0]= - Q_2,\quad [ Q_2, Q_{-1}]= -15 L_1, \nonumber \\
 [ Q_2, Q_{-2}]&= 30 Q_0 + 60 L_0,  \quad [ Q_2, Q_{-3}]= 10 Q_{-1} - 15 L_{-1},\quad [ Q_1, Q_0]= 3 L_1 - Q_1,\quad   \nonumber \\ 
 [ Q_1, Q_{-1}]&= -3 L_0 - 3 Q_0,\quad [ Q_1, Q_{-2}]= 15 L_{-1},\quad [ Q_1, Q_{-3}]= Q_{-2}, \quad [ Q_0, Q_{-1}]= - Q_{-1} - 3 L_{-1}   \nonumber \\
 [ Q_0, Q_{-2}]&= - Q_{-2},\quad [ Q_0, Q_{-3}]= Q_{-3}, \quad
 [Q_{-1}, Q_{-2}]= -6 Q_{-3},\quad  [Q_{-1}, Q_{-3}]=0,\quad [Q_{-2}, Q_{-3}]=0  \nonumber
 \end{align}

\subsection{$\mathfrak{so}(5) \supset \mathfrak{so}(3)$ chain: Poisson setting}

In this case, we consider the following coordinates: $\boldsymbol{x}:=\{\ell_{0}, \ell_1, \ell_{-1},q_3, q_2, q_1,q_0,q_{-1},q_{-2},q_{-3}\}$ and restrict to the commutant related to the subalgebra spanned by: $\{\ell_{0}, \ell_1, \ell_{-1}\}$. By applying the usual procedure, we get two polynomials of degree two (central elements):
\begin{equation}
b_1:=\ell_0^2+\ell_1 \ell_{-1},\quad b_2:=\frac{5}{3}q_1 q_{-1}-\frac{5}{2}q_0^2-\frac{1}{6}q_2 q_{-2}+q_3 q_{-3}.
\end{equation}
No polynomials of degree three arise. At degree four, we get seven polynomials together with three functional relations and restrict to four: \begin{align}
D_1&:=4 \ell_0^3 q_0 + \ell_1 \ell_{-1}^2 q_1 - \ell_{-1}^3 q_3 - \ell_1^2 \ell_{-1} q_{-1} + 
\ell_0^2 (-4 \ell_{-1} q_1 + 4 \ell_1 q_{-1}) + 
\ell_0 (-6 l_1 l_{-1} q_0 + \ell_{-1}^2 q_2 + \ell_1^2 q_{-2}) + \ell_1^3 q_{-3} \nonumber  \\
D_2&:=\ell_{-1}^2 (q_1^2 - q_0 q_2 + q_3 q_{-1}) + 
\ell_0 \ell_{-1} (-2 q_0 q_1 + q_2 q_{-1} - q_3 q_{-2}) + 
\ell_1^2 (q_{-1}^2 - q_0 q_{-2} + q_1 q_{-3}) \nonumber \\
&\hskip 0.25cm + \ell_0 \ell_1 (2 q_0 q_{-1} - q_1 q_{-2} + q_2 q_{-3}) + \ell_1 \ell_{-1} (-2 q_0^2 + q_1 q_{-1} - q_3 q_{-3}) + 
2 \ell_0^2 (2 q_0^2 - q_1 q_{-1} + q_3 q_{-3}) \nonumber\\
D_3&:=\ell_{-1} (-6 q_0^2 q_1 + 4 q_1^2 q_{-1}+ q_0 q_2 q_{-1} - 8 q_3 q_{-1}^2 - q_1 q_2 q_{-2} + 9 q_0 q_3 q_{-2} + q_2^2 q_{-3} - 12 q_1 q_3 q_{-3}) \nonumber \\
&\hskip 0.25cm + \ell_1 (6 q_0^2 q_{-1} - 4 q_1 q_{-1}^2 - q_0 q_1 q_{-2} + q_2 q_{-1} q_{-2} - q_3 q_{-2}^2 + 8 q_1^2 q_{-3} - 9 q_0 q_2 q_{-3} + 12 q_3 q_{-1} q_{-3}) \nonumber \\
&  \hskip 0.25cm + 2 \ell_0 (18 q_0^3 + 3 q_2 q_{-1}^2 + 3 q_1^2 q_{-2} - q_3 q_{-1} q_{-2} - q_1 q_2 q_{-3} + q_0 (-17 q_1 q_{-1} - 2 q_2 q_{-2} + 9 q_3 q_{-3})) \nonumber \\
D_4&:=81 q_0^4 + 16 q_1^2 q_{-1}^2 - 48 q_3 q_{-1}^3 - 8 q_1 q_2 q_{-1} q_{-2} + q_2^2 q_{-2}^2 - 12 q_1 q_3 q_{-2}^2 - 
12 (4 q_1^3 + q_2^2 q_{-1} - 12 q_1 q_3 q_{-1}) q_{-3} \nonumber \\
& \hskip 0.25 cm- 
18 q_0^2 (6 q_1 q_{-1} + q_2 q_{-2} + 18 q_3 q_{-3}) + 
24 q_0 ((q_1^2 + 3 q_3 q_{-1}) q_{-2} + q_2 (q_{-1}^2 + 3 q_1 q_{-3})) 
\end{align}
Observing that the degree-four polynomial $D_1+\frac{1}{3}D_2-\frac{1}{9}D_3+\frac{1}{108}D_4$ is central, i.e. $\{D_1+\frac{1}{3}D_2-\frac{1}{9}D_3+\frac{1}{108}D_4, \cdot\}=0$. \noindent We therefore restrict the degree four polynomials to $\{D_1, D_2, D_3\} \cup  \{d_1\}$, where $d_1:=D_1+\frac{1}{3}D_2-\frac{1}{9}D_3+\frac{1}{108}D_4$. In addition $\{d_1, \cdot\}=0$. No polynomials of degree five arise. At degree six, we get $17$ polynomials together with $12$ functional relations. This restricts us to $5$ polynomials.  We present only their leading and lowest terms in terms of ordering, which provide information to determine the complete shape of the polynomials,  i.e.: 
\[ F_1:= 12 \ell_{1}^4 q_{-1} q_{-3} -\frac{1}{64} \ell_{1}^4 q_{-2}^2 + l.o.t. ,\quad  F_2:=  6 \ell_{1}^3 q_0 q_{-1} q_{-2}-4 \ell_{1}^3 q_{-1}^3 + l.o.t.  \]
\[  F_3:=  36 \ell_{1}^2 q_1 q_{-1}^3-54 \ell_{1}^2 q_0^2 q_{-1}^2 + l.o.t , \quad  F_4:= 576 \ell_{1} q_0^2 q_1 q_{-1}^2-432 \ell_{1} q_0^4 q_{-1} + l.o.t. ,\]
\[ F_5:= 39366 q_0^4 q_{1} q_{-1}-19683 q_0^6 + l.o.t. \]
Higher order polynomial can be obtained via search in terms of degree and Poisson bracket of already defined polynomials, thus obtaining: 
\begin{align}
 G_1 &= 3 \ell_1^4 q_2 q_{-3}^2-5 \ell_1^4 q_1 q_{-2} q_{-3}+6 \ell_1^4 q_0 q_{-1} q_{-3}+2 \ell_1^4 q_0 q_{-2}^2-2 \ell_1^4 q_{-1}^2 q_{-2} + l.o.t., \\
G_2&= \ell_!^3 q_3 q_{-2}^3 - \ell_1^3 q_2 q_{-1} q_{-2}^2 + \ell_1^3 q_0 q_1 q_{-2}^2+ 4 \ell_1^3 q_1 q_{-1}^2 q_{-2} -6 \ell_1^3 q_0^2 q_{-1}q_{-2} + l.o.t. \nonumber \\
 G_3&= -12 \ell_1^2 q_2 q_{-1}^4  + 60 \ell_1^2 q_0 q_1 q_{-1}^3 -60 \ell_1^2 q_0^3 q_{-1}^2 + l.o.t. \nonumber 
\end{align}
\begin{align} 
H_1&= -16560 \ell_1 q_0^3 q_2 q_{-1}^3  + 49440 \ell_1 q_0^2 q_1^2 q_{-1}^3 -53460 \ell_1 q_0^4 q_1 q_{-1}^2 + 26730 \ell_1 q_0^6 q_{-1} + l.o.t. \nonumber \\
 I_1&= -864 \ell_1^2 q_0^2 q_2 q_{-1}^4 -2880 \ell_1^2 q_0 q_1^2 q_{-1}^4 + 7200 \ell_1^2 q_0^3 q_1 q_{-1}^3 -4320 \ell_1^2 q_0^5 q_{-1}^2 + l.o.t. \nonumber \\
I_2 &= 136080 \ell_1^2 q_0^2 q_2 q_{-1}^4+ 453600 \ell_1^2 q_0 q_1^2 q_{-1}^4 -1134000 \ell_1^2 q_0^3 q_1 q_{-1}^3 + 680400 \ell_1^2 q_0^5 q_{-1}^2 + l.o.t. \nonumber \\
 I_3&= 144 \ell_1^4 q_0^3 q_{-2}^2 + 32 \ell_1^4 q_1 q_{-1}^3 q_{-2} -48 \ell_1^4 q_0^2 q_{-1}^2 q_{-2}  + l.o.t. \nonumber \\
I_4&= 72 \ell_1^3  q_2 q_{-1}^5 -360 \ell_1^3 q_0 q_1 q_{-1}^4 + 360 \ell_1^3 q_0^3 q_{-1}^3 + l.o.t. \nonumber \\
 I_5&= 504 \ell_1^2 q_0^2 q_2 q_{-1}^4+ 1680 \ell_1^2 q_0 q_1^2 q_{-1}^4 -4200 \ell_1^2 q_0^3 q_1 q_{-1}^3 + 2520 \ell_1^2 q_0^5 q_{-1}^2 + l.o.t. \nonumber \\
L_1&= -51660 \ell_1^2 q_0^4 q_2 q_{-1}^4 + 562800 \ell_1^2 q_0^3 q_1^2 q_{-1}^4 -623700 \ell_1^2 q_0^5 q_1 q_{-1}^3 + 267300 \ell_1^2 q_0^7 q_{-1}^2 + l.o.t. \nonumber \\
L_2&= 673920 \ell_1 q_0^5 q_3 q_{-1}^4 -475200 \ell_1 q_0^4 q_1 q_2 q_{-1}^4 + 348000 \ell_1 q_0^3 q_1^3 q_{-1}^4 + 181440 \ell_1 q_0^6 q_2 q_{-1}^3 -151200 \ell_1 q_0^5 q_1^2 q_{-1}^3 + l.o.t. \nonumber 
\end{align}
with the grading
\begin{equation}
 \mathcal{G}\left(b_1\right)= (2,0),\quad  \mathcal{G}\left(b_2\right)= (0,2),\quad  \mathcal{G}\left(D_1\right)= (3,1),\quad  \mathcal{G}\left(D_2\right)= (2,2),\quad
  \mathcal{G}\left(D_3\right)= (1,3)
\end{equation}
\[    \mathcal{G}\left(D_4\right)= (0,4),\quad  \mathcal{G}\left(F_1\right)= (4,2),  \quad\mathcal{G}\left(F_2\right)= (3,3), \quad \mathcal{G}\left(F_3\right)= (2,4), \quad 
 \mathcal{G}\left(F_4\right)= (1,5) \]
 \[  \mathcal{G}\left(F_5\right)= (0,6) ,\quad  \mathcal{G}\left(G_1\right)= (4,3), \quad  \mathcal{G}\left(G_2\right)= (3,4) ,\quad  \mathcal{G}\left(G_3\right)= (2,5), \quad  \mathcal{G}\left(H_1\right)= (1,7)  \]
 \[   \mathcal{G}\left(I_1\right)= (4,5) + (2,7)+(3.6),\quad   \mathcal{G}\left(I_2\right)= (3,6)+ (2,7) ,\quad  \mathcal{G}\left(I_3\right)= (6,3)+ (5,4) + (4,5)   \]
\[    \mathcal{G}\left(I_4\right)= (5,4)+ (4,5) + (3,6) ,\quad  \mathcal{G}\left(I_5\right)= (4,5) + (3,6) + (2,7) ,\quad   \mathcal{G}\left(I_6\right)= (3,6)+ (2,7) + (1,8) ,  \]
\[   \mathcal{G}\left(I_7\right)= (2,7)+ (1,8) ,\quad    \mathcal{G}\left(L_1\right)= (4,7)+ (2,9) + (3,8),\quad  \mathcal{G}\left(L_2\right)= (3,8)+(2,9) + (1,10)  \]

\noindent At this level, we are looking for the polynomial algebra with three central elements $\{b_1, b_2, d_1\}$ generated by the $22$ polynomials $\{D_1, D_2, D_3, D_4, F_1,F_2,F_3,F_4,F_5, G_1,G_2,G_3,H_1,I_1, I_2, I_3, I_4, I_5, I_6, I_7,L_1,L_2\}$. All brackets of the form $\{D_i,F_j\}$ close linearly in terms of the $I_k$'s and $G_k$'s with some structure constants that depend on $b_k$'s. This again leads to defining another set of missing label operators from a linear combination of $D_i$'s and $F_j$'s
$\{ p_1^{(2)},  p_2^{(2)},  p_1^{(4)},  p_2^{(4)},  p_3^{(4)},  p_4^{(4)},   p_1^{(5)},   p_2^{(5)},   p_3^{(5)},  p_4^{(5)},  p_5^{(5)},  p_1^{(7)},  p_2^{(7)},  p_3^{(7)},  p_1^{(8)},   p_1^{(9)},  p_2^{(9)},$  $p_3^{(9)},  p_4^{(9)},  p_5^{(9)},  p_6^{(9)},   p_7^{(9)} , p_{1}^{(12)}, p_{2}^{(12)} \}$. We then search for closure under the bracket. Here we list some of the lower degree relations:\begin{equation}
\left\{p_{m_1}^{(i)},p_{m_2}^{(j)}\right\}= \sum_{2\alpha_1 + \ldots +  12 \alpha_{24}= i +j -1}\gamma^{\alpha_1,\ldots,\alpha_{24}}_{m_1,m_2} \quad ( p_1^{(2)}   )^{\alpha_1} \quad  \times \ldots \times  \quad  (   p_{2}^{(12)}  )^{\alpha_{24}} 
\end{equation} 
More explicitly, we get the following relations:
\begin{equation}
\{D_1,D_2\}=G_1,\quad \{D_1,D_3\}=3G_1+G_2,\quad
\{D_1,D_4\}=-12G_2
\end{equation}
\begin{equation}
\{D_2,D_3\}=9 G_1+3 G_2+G_3,\quad \{D_2,D_4\}=36 G_2+12 G_3,\quad \{D_3,D_4\}=-36 G_3
\end{equation}
All $\{D_i,F_j\}$ close in terms of $I_i$'s, $b_i$'s and $G_i$'s. Here the purpose is not to provide a complete set of relations, but to provide some of the relations which then allow to construct missing label operators,  i.e.: 
\begin{align}
\{D_2,F_4\}&=I_1,\quad 
\{D_2,F_5\}=I_2,\quad
\{D_3,F_1\}=I_3,\quad
\{D_3,F_2\}=I_4\\
\{D_3,F_3\}&=I_5,\quad
\{D_3,F_4\}=I_6,\quad
\{D_3,F_5\}=I_7 \, .
\end{align}  One example of less trivial relations among the generators is:
\begin{align}
\{D_1,F_1\}&=-\frac{304}{28143}I_1+\frac{28}{422145}I_2+\frac{2}{15} I_3-\frac{4}{45}I_4+\frac{293}{281430}I_5 -\frac{178}{140715}I_6+\frac{89}{75986100}I_7\nonumber \\
& \hskip 0.4cm-\frac{123578}{15635}b_2 G_1-\left(\frac{32 }{3}b_1 +\frac{2866}{885}b_2\right) G_2-\left(\frac{154528}{46905}b_1 +\frac{30272}{46905} b_2\right) G_3\\ \nonumber
\end{align}
This points out that certain combinations of $D_i$'s and $G_i$'s or $D_i$'s and $F_i$'s would allow further operators to commute and be added to the set of labeled operators formed by the commuting elements of the Abelian Poisson subalgebra. This will be done in a more comprehensive study of the problem elsewhere.

\section{Conclusion}

In this paper, we presented a short review of an approach for studying subalgebra chains $\mathfrak{g} \supset \mathfrak{g}'$. This approach consists in classifying the $\mathfrak{g}^{\prime}$-invariant polynomials and, in particular, the indecomposable ones. Then, in a second step, the indecomposable polynomials are used to identify a related algebraic structure, either a polynomial algebra or in the dual space a polynomial Poisson algebra. The approach uses the notion of commutant in the symmetric algebra and relies on variables of the dual space to obtain indecomposable polynomials. We reviewed the chains associated with  $\mathfrak{su}(3) \supset \mathfrak{so}(3)$, $\mathfrak{so}(5) \supset \mathfrak{su}(2) \times \mathfrak{u}(1)$,  $\mathfrak{su}(4) \supset \mathfrak{su}(2) \times \mathfrak{su}(2)$. We also provided new results and insights into the Surfon subalgebra chain $\mathfrak{so}(5) \supset \mathfrak{so}(3)$. We provided an analysis of the grading of the different polynomials. Most of the labeling problems studied in the literature provide the labeling operators of lowest possible order. However, the precise knowledge of the algebraic structure of $C_{U(\frak{g})}\left(\frak{g}^{\prime}\right)$ (or its analytical counterpart $C_{S(\frak{g})}\left(\frak{g}^{\prime}\right)$) allows us to establish the most general possible labeling operators and connecting description of the physical problem based on different sets. We plan to extend results to more general chains in the future.

\section*{Acknowledgement}
This work was partially supported by the Future Fellowship FT180100099 and the Discovery Project DP190101529 from the Australian Research Council. RCS  acknowledges financial support by the research grant PID2023-148373NB-I00 funded by MCIN/AEI/10.13039/501100011033/FEDER,UE. DL~has been partially funded by MUR - Dipartimento di Eccellenza 2023-2027, codice CUP G43C22004580005 - codice progetto: DECC23$\_$012$\_$DIP and partially supported by INFN-CSN4 (Commissione Scientifica Nazionale 4 - Fisica Teorica), MMNLP project. DL is a member of GNFM, INdAM.


\end{document}